# Polarization, Partisanship and Junk News Consumption over Social Media in the US

Vidya Narayanan
Oxford University
vidya.narayanan@oii.ox.ac.uk
@vidunarayanan

Vlad Barash
Graphika
vlad.barash@graphika.com
@vlad43210

John Kelly
Graphika
john.kelly@graphika.com
@apidictionist

Bence Kollanyi
Oxford University
bence.kollanyi@oii.ox.ac.uk
@bencekollanyi

Lisa-Maria Neudert
Oxford University
lisa-maria.neudert@oii.ox.ac.uk
@lmneudert

Philip N. Howard
Oxford University
philip.howard@oii.ox.ac.uk
@pnhoward

**ABSTRACT**
*What kinds of social media users read junk news? We examine the distribution of the most significant sources of junk news in the three months before President Donald Trump's first State of the Union Address. Drawing on a list of sources that consistently publish political news and information that is extremist, sensationalist, conspiratorial, masked commentary, fake news and other forms of junk news, we find that the distribution of such content is unevenly spread across the ideological spectrum. We demonstrate that (1) on Twitter, a network of Trump supporters shares the widest range of known junk news sources and circulates more junk news than all the other groups put together; (2) on Facebook, extreme hard right pages—distinct from Republican pages—share the widest range of known junk news sources and circulate more junk news than all the other audiences put together; (3) on average, the audiences for junk news on Twitter share a wider range of known junk news sources than audiences on Facebook's public pages.*

## POLARIZATION ON SOCIAL MEDIA

Social media has become an important source of news and information in the United States. An increasing number of users consider platforms such as Twitter and Facebook a source of news. At important moments of political and military crises, social media users not only share substantial amounts of professional news, but also share extremist, sensationalist, conspiratorial, masked commentary, fake news and other forms of junk news.[1,2]

News on social media also reaches users indirectly, when they browse social media for other purposes. With more than 2 billion monthly active users, Facebook is the most popular social media network. The *Reuters Digital News Report 2017* finds that 71% of US respondents are on Facebook, with 48% of US respondents using it for news.[3]

Given the central role that social media play in public life, these platforms have become a target for propaganda campaigns and information operations. In its review of the recent US elections, Twitter found that more than 50,000 automated accounts were linked to Russia.[4] Facebook has revealed that content from the Russian Internet Research Agency has reached 126 million US citizens before the 2016 presidential election.[5] Adding to reports about foreign influence campaigns, there is increasing evidence of a rise in polarization in the US news landscape in response to the 2016 election. Trust in news is strikingly divided across ideological lines, and an ecosystem of alternative news is flourishing, fueled by extremist, sensationalist, conspiratorial, masked commentary, fake news and other forms of junk news. At the same time, legacy publishers like the New York Times and the Washington Post have reported an increase in subscriptions.

Social media algorithms can be purposefully used to distribute polarizing political content and misinformation. Pariser's claim is that filter bubble effects—highly personalized algorithms that select what information to show in news feeds based on user preferences and behavior—have polarized public life.[6] Vicario et al. find that misinformation on social media spreads among homogeneous and polarized groups.[7] In January 2018, Facebook announced changes to its algorithm to prioritize trustworthy news, responding to ongoing public debate as to whether its algorithms promote junk content.[8] Consequently, social polarization is a driver—just as much as it may be a result—of polarized social media news consumption patterns.

In this study, we present a three-month study of junk news and political polarization among groups of US Twitter and Facebook users. In particular, we examine the distribution of posts and comments on public pages that contain links to junk news sources, across the political spectrum in the US. We then map the influence of central sources of junk political news and information that regularly publish content on hot button issues in the US. In particular, we consider patterns of interaction between accounts that have (i) shared junk news, (ii) and that have engaged with users who disseminate large amounts of misinformation about major political issues.

## SOCIAL NETWORK MAPPING

Visualizing social network data is a powerful way of understanding how people share information and associate with one another. By using selected keywords, seed accounts, and known links to particular content, it is possible to construct large network visualizations. The underlying networks of these visualizations can then be examined to find communities of accounts and clusters of association. These clusters of accounts and content can then be coded with political attributes based on knowledge of account history, content type, association metrics and social interaction between accounts.

These social network maps provide insight into both social structure and flow of information. In this study, we use the Graphika visualization suite to map and code accounts that are associated with prominent political accounts, topics, political affiliations, and geographical areas. Social network mapping also allows us to catalogue users and content, and generate both descriptive statistics and statistical models that explain changes in network structure and therefore things like information flow over time.

Social network maps comprise nodes representing the individual accounts, which are connected to other nodes in the map via social relationships. A Fruchterman–Reingold visualization algorithm can be used to represent the patterns of connection between these nodes.[9] It arranges the nodes in a visualization through a centrifugal force that pushes nodes to the edge and a cohesive force that pulls strongly connected nodes together. This mapping process produces focused "segments" of users who share very similar and specific kinds of content with each other. Segments that share some content with each other are aggregated into "groups".

The nodes in a network may all belong to a group with a shared pattern of interests. These groups can be constructed from a number of geographically, culturally, or socially similar segments. For example, segments of House Democrats, Democratic Party, Left-leaning NGOs, Liberal and anti-GOP pages, and Liberal Memes could be collectively labeled as a "Democratic Party Group". This method of segmenting users, coding groups, and generating broad observations about association is an iterative process drawing on qualitative, quantitative and computational methods. These are run many times over a period of time to identify stable and consistent communities in a network of social media users.

To create a map of segments and groups, we use a bipartite graph to provide a structural similarity metric between nodes in the map, which is used in combination with a clustering algorithm to segment the map into distinct communities. For this study, hierarchical agglomerative clustering was used to automatically generate segments and groups from sampled data (see online supplement for details).

Different social media platforms have their own unique attributes that are effective in identifying communities that persist over time. For instance, clustering Twitter users by following and follower relationships yields much more stable communities than clustering by mention or retweet relationship. Likewise, clustering Facebook users by the "like" relationship yields similarly stable results. Therefore, for this study, we have used these attributes to generate maps of stable clusters on Twitter and Facebook.

The outputs of this clustering algorithm have been extensively tested by others in studies of social media maps from Iran, Russia and the United States.[2,10,11] After clustering, the map-making process uses supervised machine learning techniques to generate labels for segments and groups from a training set labeled by human experts. After these labels are assigned, they are then manually verified and checked for accuracy and consistency.

**STUDY SAMPLE AND METHOD**

For this study, a seed of known propaganda websites across the political spectrum was used, drawing from a sample of 22,117,221 tweets collected during the US election, between November 1-11, 2016. (The full seed list is in the online supplement and available as a standalone spreadsheet.) We identified sources of junk news and information, based on a grounded typology. Sources of junk news deliberately publish misleading, deceptive or incorrect information purporting to be real news about politics, economics or culture. This content includes various forms of extremist, sensationalist, conspiratorial, masked commentary, fake news and other forms of junk news. For a source to be labeled as junk news it must fall in at least three of the following five domains:

- Professionalism: These outlets do not employ the standards and best practices of professional journalism. They refrain from providing clear information about real authors, editors, publishers and owners. They lack transparency, accountability, and do not publish corrections on debunked information.
- Style: These outlets use emotionally driven language with emotive expressions, hyperbole, ad hominem attacks, misleading headlines, excessive capitalization, unsafe generalizations and fallacies, moving images, graphic pictures and mobilizing memes.
- Credibility: These outlets rely on false information and conspiracy theories, which they often employ strategically. They report without consulting multiple sources and do not employ fact-checking methods. Their sources are often untrustworthy and their standards of news production lack credibility.
- Bias: Reporting in these outlets is highly biased and ideologically skewed, which is otherwise described as hyper-partisan reporting. These outlets frequently present opinion and commentary essays as news.
- Counterfeit: These outlets mimic professional news media. They counterfeit fonts, branding and stylistic content strategies. Commentary and junk content is stylistically disguised as news,

**Table 1: Size, Coverage and Consistency of US Audience Groups on Twitter**

|  | Users N | Users % | Coverage | Consistency |
|---|---|---|---|---|
| Conservative Media | 1,876 | 14 | 95 | 20 |
| Democratic Party | 576 | 4 | 11 | 0 |
| Local News | 469 | 3 | 28 | 0 |
| Mainstream | 744 | 6 | 33 | 1 |
| Other | 876 | 6 | 67 | 2 |
| Party Politics | 1,343 | 10 | 52 | 1 |
| Progressive Movement | 1,149 | 9 | 36 | 1 |
| Republican Party | 845 | 6 | 58 | 1 |
| Resistance | 3,663 | 27 | 62 | 18 |
| Trump Support | 1,936 | 14 | 96 | 55 |
| Average | 1,348 | 10 | 54 | 10 |
| Total | 13,477 | 100 | .. | .. |

**Table 2: Heterophily Index for US Audience Groups on Twitter**

| Group | Conservative Media | Democratic Party | Mainstream Media | Other | Local News | Party Politics | Progressive | Republican Party | Resistance | Trump Support |
|---|---|---|---|---|---|---|---|---|---|---|
| Conservative Media | 2.2 | 0.5 | 0.8 | 0.9 | 0.7 | 1.0 | 0.3 | 1.6 | 0.2 | 1.3 |
| Democratic Party | 0.5 | 1.6 | 1.7 | 1.3 | 1.5 | 1.6 | 1.9 | 0.9 | 1.4 | 0.4 |
| Mainstream Media | 0.8 | 1.7 | 3.3 | 1.3 | 1.0 | 1.5 | 1.5 | 0.9 | 1.2 | 0.2 |
| Other | 0.9 | 1.3 | 1.3 | 1.6 | 1.0 | 1.2 | 1.2 | 1.2 | 1.3 | 0.7 |
| Local News | 0.7 | 1.5 | 1.2 | 1.1 | 2.4 | 1.1 | 1.1 | 1.2 | 0.5 | 0.3 |
| Party Politics | 1.0 | 1.6 | 1.5 | 1.2 | 1.1 | 1.6 | 1.1 | 1.5 | 0.7 | 0.6 |
| Progressive | 0.3 | 1.9 | 1.5 | 1.2 | 1.1 | 1.1 | 1.9 | 0.7 | 1.0 | 0.2 |
| Republican Party | 1.6 | 0.9 | 0.9 | 1.2 | 1.2 | 1.5 | 0.7 | 1.5 | 0.5 | 1.4 |
| Resistance | 0.2 | 1.4 | 1.2 | 1.3 | 0.5 | 0.7 | 1.0 | 0.5 | 2.0 | 0.0 |
| Trump Support | 1.3 | 0.4 | 0.2 | 0.7 | 0.3 | 0.6 | 0.2 | 1.4 | 0.0 | 4.0 |

**Figure 1: US Audience Groups on Twitter**

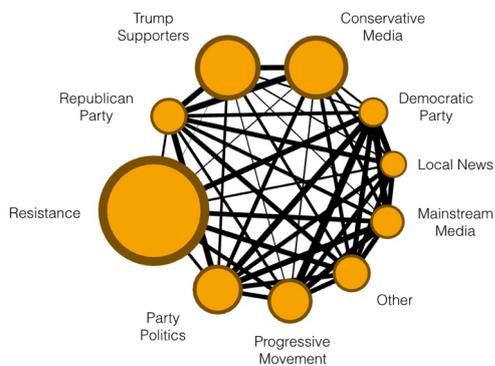

*Source: Authors' calculations from data sampled 20/10/17-20/01/2018. Note: Groups are determined through network association and our interpretation of the kinds of content these users distribute. This is a basic visualization, see online supplement for a full visualization.*

with references to news agencies, and credible sources, and headlines written in a news tone, with bylines, date, time and location stamps.

Sources of junk news were evaluated and re-evaluated in a rigorously iterative coding process. A team of 12 trained coders, familiar with the US political and media landscape, labeled sources of news and information based on a grounded typology. The Krippendorff's alpha value for inter-coder reliability among three executive coders, who developed the grounded typology, was 0.805. The 91 sources of political news and information, which we identified over the course of several years of research and monitoring, produce content that includes various forms of propaganda and ideologically extreme, hyper-partisan, and conspiratorial political information. We tracked how the URLs to these websites were being shared over Twitter and Facebook (see online supplement for details).

Specifically, we computed the coverage and consistency scores for each group. *Coverage* of a group refers to the percentage of all propaganda domains identified in our junk news sources list that a group posted links to. The *Consistency* of a group refers to the percentage of the total of number of links to all the propaganda domains identified in our junk news sources list, that is shared by the group. A high value for coverage shows that the group is sharing a wide range of propaganda, while a high value for consistency shows that the group is playing a key role in the spreading of such propaganda. Coverage and consistency scores were calculated from the number of links shared from the groups to the junk news sources.

**FINDING: POLARIZATION AND JUNK NEWS ON TWITTER**

Our Twitter dataset contains 13,477 Twitter users collected during a 90-day period between October 20, 2017 and January 18, 2018. To study the polarization among US audience groups on Twitter, we first identified the accounts of Democratic and Republican party members, at both state and national levels. Further, we identified Twitter accounts of members of congress from both parties. Next, we included all the followers of these accounts in our dataset. We identified a follower network of 93,711 Twitter accounts. We then reduced this sample of Twitter users to a set of well-connected accounts using a variant of k-core reduction (see online supplement for details).[12] This reduced the dataset to 13,477 Twitter users. Finally, we collected all Twitter users followed by any account in the reduced set of Twitter users, in order to segment this set into communities of interest.

We used Twitter's REST API to collect publicly available data for our analysis. Twitter's REST API provides data on a) who follows whom on Twitter (100% of all data), and b) recent tweets for each user (up to 3,200 tweets per user in reverse chronological order).

Twitter's APIs give access only to public data and do not provide any information about suspended accounts or users who set their accounts private. The latter limitation is not a concern here, given that 100% of Twitter users in this study have public accounts.[13]

We were able to group our sample of 13,477 user accounts into 10 groups of affiliation. The groups emerged through network association, and by interpretation of the kinds of content these users distributed and indicated as a "favorite". Table 1 identifies the main groupings of US Twitter users sampled, as labelled by our iterative machine-learning process and expert manual review.

From Table 1, we see that the Trump Support Group has a coverage of 96%, indicating that those pages share the widest range of junk sources on Twitter. This is followed by the Conservative Media Group, with a coverage of 95%. We also see from Table 1 that the Trump Support group, with a consistency score of 55%, contributes more to the spreading of junk news, compared to all other groups put together.

Next, we next calculated a heterophily score for each combination of group pairings. This is a measure of the connections between groups in a network, where a ratio is calculated of the actual ties between two groups, compared with the expected number of ties between them, if all the ties in the network were distributed evenly. We calculate ties for groups on Twitter from follower accounts and accounts followed, and Facebook ties from page likes. The natural log of the ratios is then taken along with a zero correction to create a balanced index. A high heterophily score between groups indicates more connections between the two groups. A high heterophily score for a group to itself indicates a high number of within-group connections. It is important to note however that these scores indicate only first order (direct) connections between groups, and not second, third, or higher-order (indirect) connections. These values are shown in Table 2.

From Table 2, we see that the Democratic Party Group and the Mainstream Media Group have a heterophily index of 1.7, indicating a deep connection between the two groups. A heterophily score of 1.0 would indicate a perfectly neutral level of connection between groups; less than 1.0 would indicate a lack of connection. Similarly, we see that the Republican Party Group shares a heterophily index of 1.6 with the Conservative Media Group, indicating strong interactions between them. The Democratic Party also shares a high heterophily index of 1.9 with the Progressive Movement Group, demonstrating significant interaction. The Mainstream Media Group also shares a high heterophily score with both the Progressive Movement (1.5), and the Resistance (1.2) Groups. The Republican Party and Trump Supporters share a heterophily score of 1.4, also indicating a strong connection between them.

Figure 1 is a basic visualization of the 10 groups on Twitter. The size of each group is determined by the number of Twitter accounts that belong to it (see Table 1). The connections between the groups in the figure are computed using the heterophily scores (see Table 2). The width of the line

**Table 3: Size, Coverage and Consistency of US Audience Groups on Facebook**

| | Users N | Users % | Coverage | Consistency |
|---|---|---|---|---|
| Conspiracy | 946 | 9 | 40 | 5 |
| Democratic Party | 1,144 | 11 | 40 | 12 |
| Environmental Movement | 954 | 9 | 13 | 1 |
| Hard Conservative | 815 | 8 | 91 | 58 |
| Libertarians | 209 | 2 | 34 | 4 |
| Military Guns | 397 | 4 | 45 | 4 |
| Occupy | 1,114 | 10 | 38 | 7 |
| Other Left | 673 | 6 | 6 | 2 |
| Other Non-Political | 1,688 | 16 | 13 | 2 |
| Public Health | 733 | 7 | 4 | 0 |
| Republican Party | 241 | 2 | 15 | 1 |
| Sustainable Farming | 1,144 | 11 | 19 | 2 |
| Women's Rights | 633 | 6 | 13 | 1 |
| Average | 765 | 7 | 33 | 9 |
| Total | 10,691 | 100 | .. | .. |

**Table 4: Heterophily Index for US Audience Groups on Facebook**

| Group | Conspiracy | Democratic Party | Environmental | Hard Conservative | Libertarians | Military Guns | Occupy | Other Left | Other Non-Political | Public Health | Republican Party | Sustainable Farming | Women's Rights |
|---|---|---|---|---|---|---|---|---|---|---|---|---|---|
| Conspiracy | 5.0 | 0.8 | 0.5 | 0.4 | 2.5 | 0.2 | 1.0 | 0.1 | 0.8 | 0.1 | 0.0 | 0.4 | 0.1 |
| Democratic Party | 0.8 | 5.0 | 0.6 | 0.3 | 0.6 | 0.0 | 0.8 | 1.4 | 0.7 | 0.3 | 0.7 | 0.2 | 0.7 |
| Environmental Movement | 0.5 | 0.6 | 5.7 | 0.0 | 0.0 | 0.1 | 0.4 | 0.3 | 0.7 | 0.7 | 0.8 | 1.6 | 0.2 |
| Hard Conservative | 0.4 | 0.2 | 0.0 | 9.2 | 2.2 | 1.9 | 0.0 | 0.0 | 0.3 | 0.1 | 1.8 | 0.1 | 0.0 |
| Libertarians | 2.5 | 0.6 | 0.0 | 2.2 | 26 | 0.6 | 0.3 | 0.0 | 0.2 | 0.0 | 0.3 | 0.2 | 0.0 |
| Military Guns | 0.2 | 0.2 | 0.1 | 2.0 | 0.6 | 18 | 0.0 | 0.0 | 0.6 | 0.6 | 0.8 | 0.2 | 0.1 |
| Occupy | 1.0 | 0.8 | 0.4 | 0.0 | 0.3 | 0.0 | 6.2 | 0.4 | 0.2 | 0.0 | 0.0 | 0.2 | 0.1 |
| Other Left | 0.1 | 1.4 | 0.3 | 0.0 | 0.1 | 0.0 | 0.4 | 8.6 | 0.7 | 0.6 | 0.1 | 0.1 | 0.9 |
| Other Non-Political | 0.8 | 0.7 | 0.7 | 0.3 | 0.2 | 0.6 | 0.2 | 0.7 | 3.0 | 1.3 | 0.5 | 0.8 | 0.7 |
| Public Health | 0.0 | 0.3 | 0.7 | 0.1 | 0.0 | 0.6 | 0.0 | 0.3 | 1.3 | 7.2 | 0.3 | 0.5 | 1.6 |
| Republican Party | 0.0 | 0.7 | 0.0 | 1.8 | 0.3 | 0.8 | 0.0 | 0.2 | 0.5 | 0.3 | 25 | 0.0 | 0.1 |
| Sustainable Farming | 0.4 | 0.2 | 1.6 | 0.1 | 0.2 | 0.2 | 0.2 | 0.1 | 0.8 | 0.5 | 0.0 | 5.4 | 0.2 |
| Women's Rights | 0.1 | 0.7 | 0.2 | 0.0 | 0.0 | 0.1 | 0.1 | 0.9 | 0.7 | 1.6 | 0.1 | 0.2 | 9.4 |

**Figure 2: US Audience Groups on Facebook**

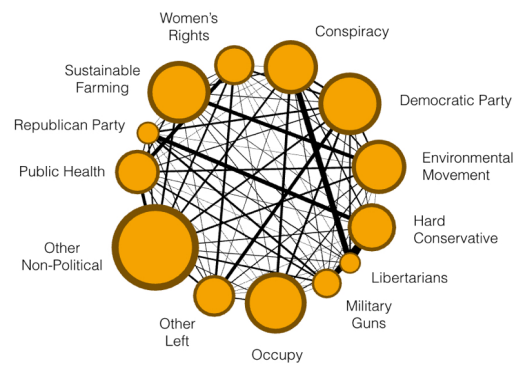

*Source: Authors' calculations from data sampled 20/10/17-20/01/2018. Note: Groups are determined through network association and our interpretation of the kinds of content these users distribute. This is a basic visualization, see online supplement for a full visualization.*

linking groups in the figure, represents the strength of connection between them.

**FINDING: POLARIZATION AND JUNK NEWS ON PUBLIC FACEBOOK PAGES**
We mapped the public Facebook pages by combining: 1) harvested Facebook public page seeds from political tweets shared during the US election and a snowball sample of the wider Facebook network around these key online interest groups; 2) a snowball sample of all the Facebook pages associated with party Twitter accounts considered for the Twitter study; 3) iteration of clear US Liberal and Conservative clusters from previous US political maps on Facebook.

This resulted in a dataset of 47,719 public Facebook pages. We then reduced this sample to a set of well-connected pages using a variant of k-core reduction (see online supplement for details) From this reduced dataset of 10,691 pages, we collected all posts from the 90 days between October 20, 2017 and January 19, 2018, using the Facebook Graph API. We extracted all URLs from posts, and analyzed the pattern of web citations across the major groupings we identified in the US news ecosystem on Facebook. Additionally, we collected the share counts for all posts containing the identified URLs from our seed list in order to measure the degree to which junk news content from various sources is shared across the Facebook network. This value includes shares that occur on private pages.

Table 3 identifies the main groupings of the US Facebook pages sampled. The Facebook groups were identified by following the same procedure that we used for the Twitter dataset.

From the coverage and consistency scores in Table 3, we see that the Hard Conservatives Group has a coverage score of 91%, followed by the Military and Guns Group at 45% and then the Conspiracy Group and Democrats Group at 40%. The Hard Conservatives Group also has a consistency score of 58%, indicating that this group has a greater share in the distribution of junk news on Facebook than all the other groups put together.

The heterophily scores for each pair of Facebook groups is shown in Table 4. We see that the heterophily score between the Conspiracy Group and almost all other groups is less than 1.0, indicating a low level of social interaction. The two key exceptions are the Libertarians Group at 2.5 and the Occupy Group at 1.0. These scores show that the Conspiracy Group is most connected to the fringes of the US political spectrum. Further, we observe that the Hard Conservative and the Libertarian Groups also interact closely with each other (heterophily score of 2.2).

Figure 2 is a basic visualization of the 13 groups on Facebook. The size of each group is determined by the number of Facebook pages that belong to it (see Table 3). The connections between the groups in the figure are computed using the heterophily scores (see Table 4). The width of the lines linking groups in the figure represents the strength of connection between them.

**CONCLUSIONS**
On Twitter, the Trump Support Group shares 95% of the junk news sites on the watch list, and accounted for 55% of junk news traffic in the sample. Other kinds of audiences shared content from these junk news sources, but at much lower levels. On Facebook, the Hard Conservative Group shares 91% of the junk news sites on the watch list, and accounted for 58% of junk news traffic in the sample. The coverage and consistency scores for Facebook and Twitter reveal some important features of these platforms when it comes to junk news circulation. The average coverage score for the major audiences of junk news on Twitter and Facebook is 54 and 33, respectively. This means that on average, groups of Twitter share 54% of the junk news watch list and groups of Facebook users share 33%.

The social networks mapped from public Twitter and Facebook data show that the junk political news and information was concentrated among Trump's supporters. The two main political parties, Democrats and Republicans, prefer different sources of political news, with limited overlap. For instance, the Democratic Party shows high levels of engagement with mainstream media sources and the Republican Party with Conservative Media Groups. On Twitter in particular, the Democratic Party have interacted closely with the Progressive Movements Group, suggesting a broad intersection of interests. On Facebook, most connections between groups conform to the partisan polarization found on Twitter. We also find close interactions between the Occupy Group and the Conspiracy Group.

**ONLINE SUPPLEMENTS AND DATA SHEETS**
Please visit comprop.oii.ox.ac.uk for additional material related to the analysis, including (1) high-resolution maps of the networks for both Twitter and Facebook, showing all accounts separated into 48 segments within the 10 groups on Twitter, and 48 segments within the 13 groups on Facebook, (2) the full list of segments and groups, (3) calculation of heterophily scores, (4) detailed explanation of the hierarchical agglomerative clustering algorithm used to create groupings, (5) the k-core reduction used to reduce the set of Twitter users, (6) a detailed description of the junk news classification methodology and, (7) a list of the junk news sites that we used for this study.

**ABOUT THE PROJECT**
The Project on Computational Propaganda (http://comprop.oii.ox.ac.uk/) involves international, and interdisciplinary, researchers in the investigation of the impact of automated scripts—computational

propaganda—on public life. *Data Memos* are designed to present quick snapshots of analysis on current events in a short format. They reflect methodological experience and considered analysis, but have not been peer-reviewed. *Working Papers* present deeper analysis and extended arguments that have been collegially reviewed and that engage with public issues. The Project's articles, book chapters and books are significant manuscripts that have been through peer review and formally published.

**ACKNOWLEDGMENTS AND DISCLOSURES**

The authors gratefully acknowledge the support of the (1) National Science Foundation, "EAGER CNS: Computational Propaganda and the Production / Detection of Bots," BIGDATA-1450193, 2014-16, Philip N. Howard, Principal Investigator; (2) the European Research Council, "Computational Propaganda: Investigating the Impact of Algorithms and Bots on Political Discourse in Europe," Proposal 648311, 2015-2020, Philip N. Howard, Principal Investigator; and (3) the Engineering and Physical Sciences Research Council (EPSRC). Project activities were approved by the University of Washington Human Subjects Committee, approval #48103-EG, and the University of Oxford's Research Ethics Committee. Any opinions, findings, and conclusions or recommendations expressed in this material are those of the authors and do not necessarily reflect the views of the National Science Foundation, the European Research Council, the Engineering and Physical Sciences Research Council, or the University of Oxford.

**Polarization, Partisanship and Junk News Consumption over Social Media in the US**

COMPROP DATA MEMO 2018.1 – Online Supplement / FEBRUARY 6, 2018

Vidya Narayanan
Oxford University
vidya.narayanan@oii.ox.ac.uk
@vidunarayanan

Vlad Barash
Graphika
vlad.barash@graphika.com
@vlad43210

John Kelly
Graphika
john.kelly@graphika.com
@apidictionist

Bence Kollanyi
Oxford University
bence.kollanyi@oii.ox.ac.uk
@bencekollanyi

Lisa-Maria Neudert
Oxford University
lisa-maria.neudert@oii.ox.ac.uk
@lmneudert

Philip N. Howard
Oxford University
philip.howard@oii.ox.ac.uk
@pnhoward

## 1. US Audience Groups on Twitter

Each node in this network (Figure 1) represents an account on Twitter. Each node belongs to both a broad group and a smaller segment within that group. A segment is a collection of nodes with a shared pattern of interest while a group is a collection of segments that are politically, culturally, or socially similar. The size of each node is proportional to the number of other nodes that follow it on Twitter. The color of each node is based on its parent segment. Figure 1 is the full color visualization of the US audience groups we found on Twitter.

**Figure 1: Full Illustration of US Audience Groups on Twitter**

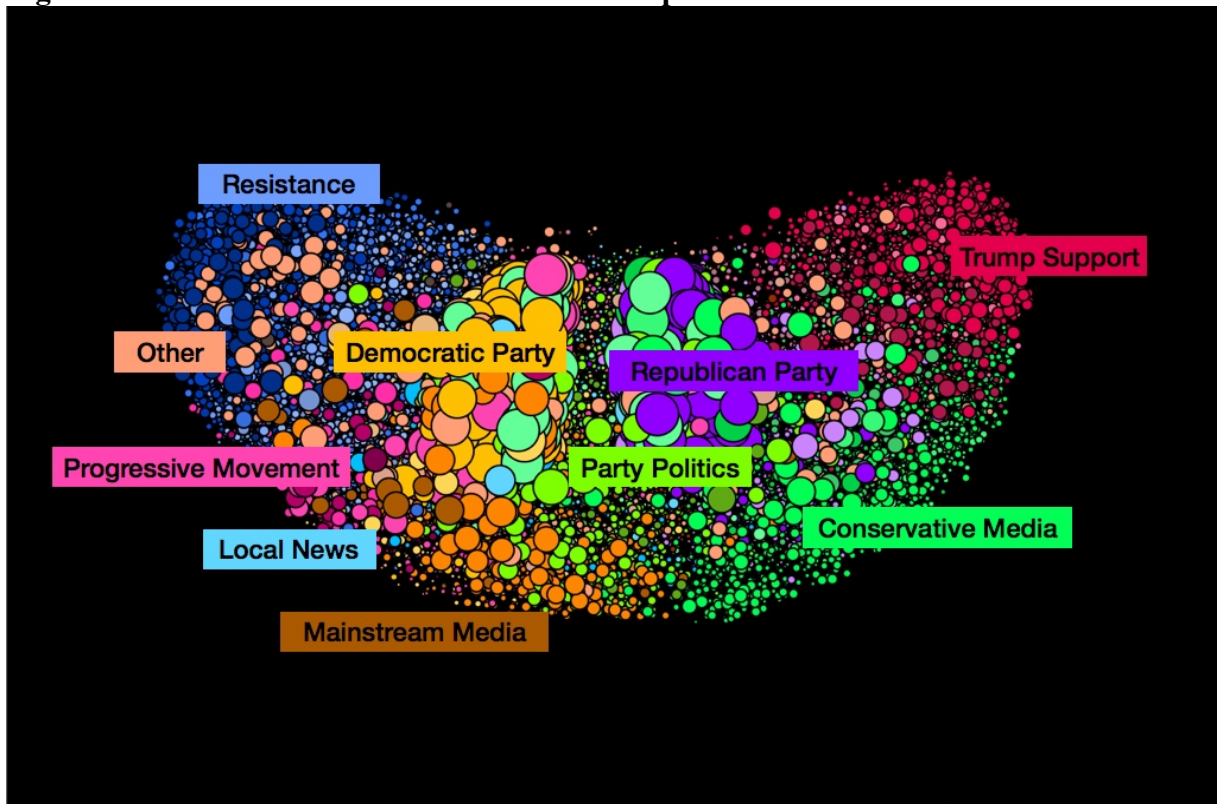

Source: Authors' calculations from data sampled 20/10/17-20/01/18. Note: Groups are determined through network association and our interpretation of the kinds of content these users distribute.

The nodes are placed within the map using a Fruchterman-Reingold visualization algorithm. This works to place nodes on the map according to two principles: first, a "centrifugal force" acts upon each node to push it to the edge of the canvas; second, a "cohesive force" acts upon every connected pair of nodes to pull them closer together. Table 1 gives a full list of Twitter groups and associated segments considered in this study.

**Table 1: Groups and Segments for the Twitter Visualization**

| Groups | Segments |
|---|---|
| Conservative Media | Conservative Blogs |
| | Conservative Truth |
| | GOP Communications |
| | Hard Right Journalists |
| | Right Wing Media |
| | Watch Dogs |
| Democratic Party | House Democrats |
| | State Democrats |
| Local News | Bay Area News |
| | Health News |
| | NYC News |
| | Texas Journalists |
| Mainstream Media | Left Wing Political Journalists |
| | News Anchors |
| Other | Brand Digital Marketing |
| | EU, Euro Organizations |
| | SMM Tech/Finance |
| Party Politics | Bi-Partisan Research |
| | GOP, Democrats Communications |
| | House – GOP |
| | US Congress |
| | US Senate |
| Progressive Movement | African American Culture and Empowerment |
| | Latino Empowerment and Non-profits |
| | Race and Grassroots |
| | Wage reform and Economic Justice |
| Republican Party | GOP |
| | Organized GOP |
| Resistance | Anti-Trump Political Analysts |
| | Anti-Trump, Human Rights, Humor |
| | Impeach Trump |
| | Pro Hilary – I'm with Her |
| | Resist – Analysts |
| | Resist – Trump |
| | Resist – Equal Rights |
| | Resist – Progressive |
| | The Resistance – Not My President |
| | Unite Democrats |
| | Women's Issues |
| Trump Support | Defend Trump/Lifestyle/Humor |
| | Patriots for Trump |
| | Re-elect Trump |
| | Trump Supporters MAGA |
| | Trump Train |
| | Trump Victory |
| | Trump Victory Party |

## 2. US Audience Groups on Facebook

Each node in this network (Figure 2) represents a public page on Facebook. The size of each node corresponds to the number of other nodes that like the page on Facebook. Each node belongs to both a broad group and a smaller segment within that group. A segment is a collection of nodes with a shared pattern of interest while a group is a collection of segments that are politically, culturally, or socially similar.

**Figure 2: Full Illustration of US Audience Groups on Facebook**

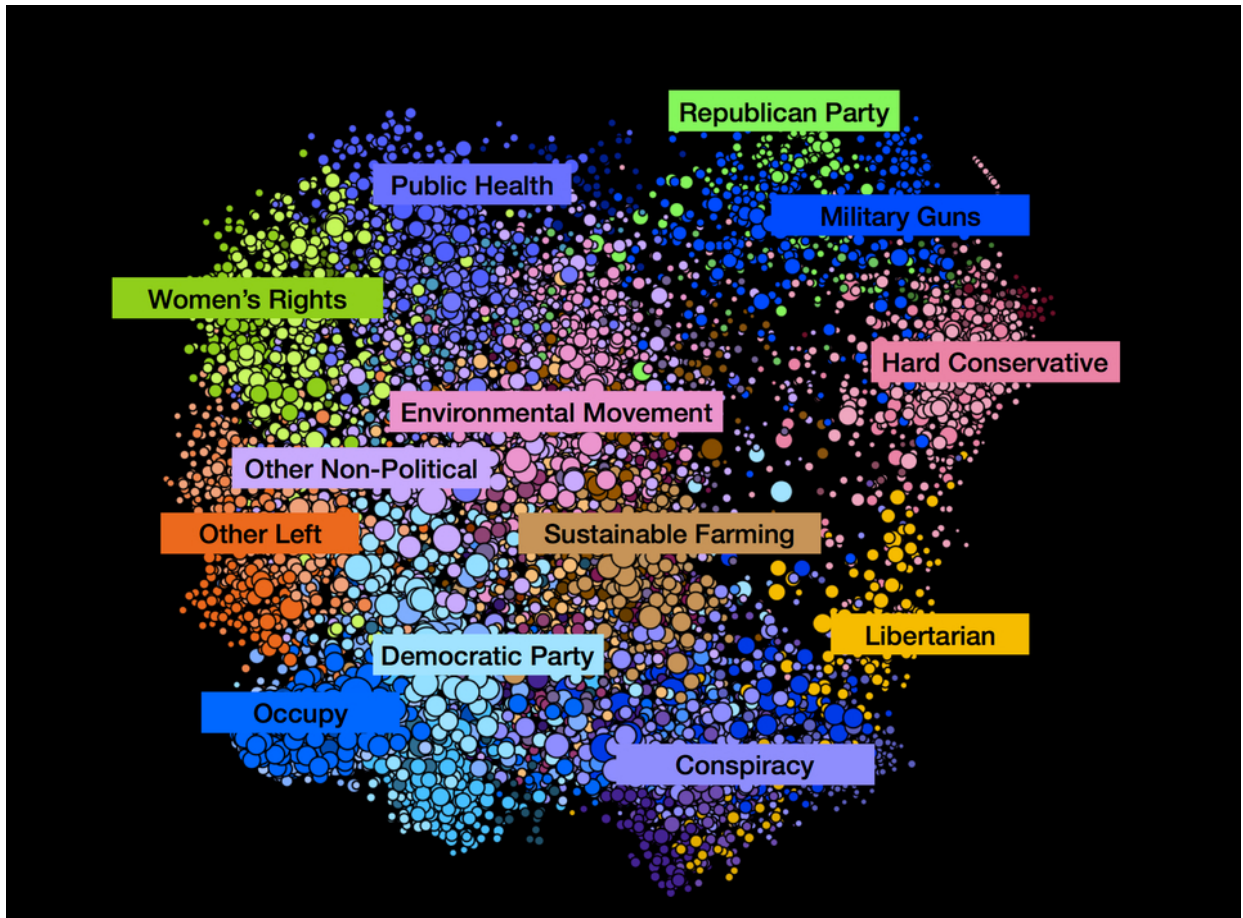

Source: Authors' calculations from data sampled 20/10/17-20/01/18. Note: Groups are determined through network association and our interpretation of the kinds of content these users distribute.

Again, a Fruchterman-Reingold visualization algorithm is used to place nodes within the map. Table 2 gives a full list of the Facebook groups and associated segments considered in this study.

**Table 2: List of Groups and Segments for the Facebook Visualization**

| Groups | Segments |
|---|---|
| Conspiracy | Conspiracy Theories |
| | Infowars |
| | International anti-media conspiracy |
| | Truthers |
| Democratic Party | Anti-GOP |
| | Atheism |
| | Democratic Humor |
| | Democratic Memes |
| | Democratic Party |
| | House Democrats |
| | Non-profit campaigns |

| Groups | Segments |
|---|---|
| Environmental Movement | Animal rights |
| | Animal, wilderness conservation |
| | Anti-fracking, eco defense |
| | Anti-GMO |
| | International climate action |
| | International Greenpeace |
| | Native American, eco resistance |
| Hard Conservatives | Anti-immigration |
| | Anti-liberal |
| | Hard Conservatives Media |
| | Patriots, Pro-Military, Militia |
| Libertarians | Anarchist |
| | Libertarian Institutes |
| Military/Guns | Guns, Guns Ownership |
| | US Army, Navy, Veterans |
| Occupy | International Anarchist |
| | International Anonymous |
| | International Occupy chapters |
| | International Occupy movement |
| | US Occupy movement |
| Other Left | ACLU |
| | Grassroots movements |
| | Immigrant Advocacy |
| | Labor reform |
| | Organized Labor, Trade Unions |
| | US and International LGBT advocacy |
| Other Non-Political | Food blogs |
| | Life coach, inspirational groups |
| | Miscellaneous Influencers |
| | Music, music clusters, SMM |
| | Parenting |
| | Spiritualism |
| | Un-clustered |
| | US and International Art, Art museums |
| Public Health | Global Health |
| | Health organizations |
| | Public initiatives |
| | US State Departments, US Consulates |
| Republican Party | GOP Congress |
| | GOP Senate |
| Sustainable Farming | Gardening sustainability |
| | Healthy eating bloggers |
| | Organic farming |
| | Organic food |
| | Small farming, self-sufficiency |
| | Survivalist |
| | Sustainable agriculture and research |
| Women's Rights | Feminist groups |
| | International women's rights |
| | Planned Parenthood |
| | Reproductive rights |

## 3. Heterophily Index

For every pairing of groups within a network map, a value of heterophily can be calculated. This is a measure of the level of connection between the groups. In order to determine this a ratio is calculated between the actual ties between two groups, compared to the expected ties between the groups if all the accounts in the map were evenly distributed. The natural log of these ratios is then taken, along with a zero correction to create a balanced index and to ensure that all values are displayed in a positive form.

$$Ratio\ of\ Ratios_T = \frac{\dfrac{Connections_{pairing}}{\sum_{all\ pairings} Connections}}{\dfrac{Connections_{pairing}}{\sum_{all\ pairings} Connections}}$$

Expression A: Ratio of Two Ratios

This heterophily index is therefore created through a ratio of two ratios. This ratio reveals whether two nodes have about the proportion of links they should have given its size. This is displayed in Expression A, where a pairing of groups is calculated as having a measure of connections in balance with its share of all the connections.

Half the distribution of possible values from this ratio of ratios ranges from 0 to 1 (corresponding to disproportionately small share of connections in a group given its size) and the other half ranges from 1 to +infinity (a disproportionately large share of connections in a group given its size). However, by taking the natural log of the ratio of ratios the index will become more balanced: from -infinity to 0 becomes less than proportionate share, and from 0 to +infinity becomes more than proportionate share. For example, take a three-group network (A, B and C). If nodes in group A have a total of ten connections, and there are ten nodes in each group, then the expected connections between A and B will be 3.33. If, in reality, the nodes in group A actually have all ten connections to nodes in group B then this connection is stronger than expected. The heterophily score for groups A and B = 10/3.33 = 3.0. The natural log of this is then taken along with a zero correction across the range of heterophily values.

A greater heterophily index indicates a denser pattern of connections between the two groups. It is important to note however that these scores indicate only first order connections, not second or third order connections.

### 4. Clustering Algorithm for Determining Groups and Segments
In order to generate segments and groups for each map it is necessary to employ a clustering algorithm. This involves first building a bipartite graph between nodes in the map and the rest of the social medium in question. This bipartite graph provides a structural similarity metric between nodes in the map. This was then used in combination with a hierarchical agglomerative clustering algorithm in order to segment a map into distinct communities. This is a 'bottom up' approach whereby each observation starts in its own cluster, and pairs of clusters are merged as one moves up the hierarchy. Twitter maps are clustered based on follower-relationships, since mention-relationships have been shown to overemphasize the news cycle and salient external events. Facebook networks are clustered based on page likes.

### 5. K-core reduction
To identify and map the 'discussion core' of the most active, connected, and influential users, we performed a k-core reduction to reduce the total collected set of Twitter users from the initial data collection into a set of well-connected accounts. This produces a maximally connected subgraph of active nodes with degree of connection at least *'k'*. This degree of connection, k, can be thought of as the number of links between each node in the graph. For example, selecting a k value of 0 for the reduction does not remove any nodes from the graph, since each node must have 0 connections or greater. Selecting a k value of 1 would remove all of the nodes that have no connections to other nodes in the graph. Selecting a k value of 2 would remove all nodes with fewer than 2 connections, and so on. A value of k was selected such that the k-core consisted of 12,413 users. This value was found to be a sufficiently large group to represent the major sets of highly active users, but not so large as to make clustering and visualization impractical.

## 6. Junk News Classification

These sources deliberately publish misleading, deceptive or incorrect information purporting to be real news about politics, economics or culture. This content includes various forms of propaganda and ideologically extreme, hyper-partisan, or conspiratorial news and information. For a source to be labelled as junk news at least three of the following five characteristics must apply:

- Professionalism: These outlets do not employ the standards and best practices of professional journalism. They refrain from providing clear information about real authors, editors, publishers and owners. They lack transparency, accountability, and do not publish corrections on debunked information.
- Style: These outlets use emotionally driven language with emotive expressions, hyperbole, ad hominem attacks, misleading headlines, excessive capitalization, unsafe generalizations and fallacies, moving images, graphic pictures and mobilizing memes.
- Credibility: These outlets rely on false information and conspiracy theories, which they often employ strategically. They report without consulting multiple sources and do not employ fact-checking methods. Their sources are often untrustworthy and their standards of news production lack credibility.
- Bias: Reporting in these outlets is highly biased and ideologically skewed, which is otherwise described as hyper-partisan reporting. These outlets frequently present opinion and commentary essays as news.
- Counterfeit: These outlets mimic professional news media. They counterfeit fonts, branding and stylistic content strategies. Commentary and junk content is stylistically disguised as news, with references to news agencies, and credible sources, and headlines written in a news tone, with bylines, date, time and location stamps.

Table 3 gives a list of the all junk news sources used for this analysis.

### Table 3: List of Junk News Sources

| Domain Name | Example URL |
|---|---|
| 100percentfedup.com | http://100percentfedup.com/wow-woman-delivers-knock-out-punch-to-michelle-obama-on-way-to-polls-yes-i-am-black-but-i-am-not-voting-democrati-am-not-on-that-plantation-video/ |
| allenbwest.com | http://www.allenbwest.com/michellejesse/bombshell-new-email-shows-pentagon-tried-to-send-help-in-benghazi-but |
| americanthinker.com | http://www.americanthinker.com/articles/2016/10/what_kind_of_genius_loses_6_billion_hillary.html |
| anonews.co | http://www.anonews.co/hillary-clinton-exposed/ |
| barenakedislam.com | http://www.barenakedislam.com/2016/11/03/fbi-sources-believe-clinton-foundation-case-is-likely-moving-toward-an-indictment/ |
| beforeitsnews.com | http://beforeitsnews.com/politics/2016/11/breaking-wikileaks-to-drop-hillarys-33k-deleted-emails-tomorrow-video-2855077.html |

| Domain Name | Example URL |
| --- | --- |
| bipartisanreport.com | http://bipartisanreport.com/2016/09/09/just-in-barbara-bush-verbally-dissects-any-woman-who-votes-for-donald-trump-video/ |
| bizpacreview.com | http://www.bizpacreview.com/2016/11/04/refuse-pawn-huffpo-writer-latino-activist-recants-hillary-support-powerful-op-ed-408408 |
| bredred.com | http://bredred.com/republicans-have-cast-17000-more-votes-than-dems-in-fl/ |
| breitbart.com | http://www.breitbart.com/video/2016/11/06/nate-silver-clinton-one-state-away-losing-electoral-college/ |
| campusreform.org | https://www.campusreform.org/?id=8352 |
| centerforsecuritypolicy.org | https://www.centerforsecuritypolicy.org/civilization-jihad-reader-series/ |
| clintonemail.com | http://clintonemail.com |
| cnsnews.com | http://cnsnews.com/news/article/susan-jones/americans-not-labor-force-participation-rate-risesdrops |
| commonsense-conservative.org | http://commonsense-conservative.org/?p=1761 |
| concealncarry.stfi.re | http://concealncarry.stfi.re/forum/forum/general-discussions/2nd-amendent-discussions/778-why-i-carry-a-gun-by-urban-carry-holsters?sf=eyenxnw |
| conservativedailypost.com | https://conservativedailypost.com/breaking-fbi-confirms-evidence-of-huge-underground-clinton-sex-network/ |
| conservativeoutfitters.com | https://www.conservativeoutfitters.com/blogs/news/92961857-john-kasich-took-202-700-from-george-soros |
| conservativeread.com | http://conservativeread.com/vote-like-your-guns-depend-on-it-they-do-trumppence16/#.wcibkuhgady.twitter |
| conservativereview.com | https://www.conservativereview.com/commentary/2016/11/this-new-ted-cruz-video-is-the-last-obamacare-fact-check-you-will-ever-need |
| conservativetribune.com | https://conservativetribune.com/wikileaks-reveals-what-bill-hiding/?utm_source=twitter&utm_medium=conserv_tribune |
| constitution.com | http://constitution.com/iran-reports-u-s-supplying-isis/ |
| crooksandliars.com | http://crooksandliars.com/2016/11/latinos-nevada-and-florida-are-building |
| dailycaller.com | http://dailycaller.com/2016/10/30/hillary-clinton-knew-she-was-helping-islamists-move-into-power-in-libya/?utm_campaign=atdailycaller&utm_source=twitter&utm_medium=social |
| dailynewsbin.com | http://www.dailynewsbin.com/?p=26577 |
| dangerandplay.com | http://www.dangerandplay.com/2016/11/02/how-a-mindset-expert-who-hates-politics-got-involved-with-maga3x/ |
| dcclothesline.com | http://www.dcclothesline.com/2016/11/02/confirmed-us-intel-operatives-leaked-clinton-campaign-emails-not-russia/ |
| deepstatenation.com | http://deepstatenation.com/2016/11/oklahoma-republican-calls-for-hillary-clinton-to-be-shot-to-death-screenshots/ |
| dennismichaellynch.com | http://dennismichaellynch.com/podestacooking1/ |
| donaldtrumpnews.co | http://donaldtrumpnews.co/news/biggest-star-comes-trump-matthew-mcconaughey-votes-trump/ |
| drudgereport.com | http://www.drudgereport.com/ |
| endingthefed.com | http://endingthefed.com/breaking-bombshell-wikileaks-exposes-democrats-fake-trump-groping-plot.html |
| eutimes.net | http://www.eutimes.net/2016/10/hillary-clintons-sudden-move-of-1-8-billion-to-qatar-central-bank-stuns-financial-world/ |
| floppingaces.net | http://www.floppingaces.net/2016/10/09/enabler-hillarys-actions-speak-louder-than-trumps-words-guest-post/?platform=hootsuite |
| freebeacon.com | http://freebeacon.com/politics/mccaskill-fbi-probe-exists-legitimate-question-whether-clinton-broke-law/ |
| frontpagemag.com | http://www.frontpagemag.com/fpm/264540/seven-clinton-policy-priorities-would-devastate-john-perazzo |
| gotnews.com | http://gotnews.com/breaking-ex-apprentice-summerzervos-paid-500000-gloriaallred-accuse-trump-deal-went-others/ |
| hannity.com | http://www.hannity.com/articles/hanpr-election-493995/watch-undercover-journalist-in-full-burka-15277316/ |
| hotair.com | http://hotair.com/archives/2016/11/10/newt-has-no-time-for-whiny-sniveling-negative-nevertrump-cowards/ |
| hotpagenews.com | http://hotpagenews.com/r/162858 |

| Domain Name | Example URL |
| --- | --- |
| infowars.com | http://www.infowars.com/spirit-cooking-clinton-campaign-chairman-invited-to-bizarre-satanic-performance/ |
| inquisitr.com | https://www.inquisitr.com/3680558/new-wikileaks-emails-suggest-bernie-sanders-was-leveraged-into-endorsing-clinton/ |
| joeforamerica.com | http://joeforamerica.com/?p=55358 |
| judicialwatch.org | http://www.judicialwatch.org/press-room/press-releases/judicial-watch-court-hearing-monday-november-7-fbi-clinton-email-records-case/ |
| lawnews.tv | http://lawnews.tv/wikileaks-clinton-camp-worried-uranium-one-deal-being-investigated-hillary-sold-20-of-americas-uranium-to-russia/ |
| lifenews.com | http://www.lifenews.com/2016/10/31/black-pastor-urges-christians-hillarys-deplorables-to-vote-for-donald-trump/ |
| magafeed.com | http://magafeed.com/the_donald-uncovers-dark-connections-between-the-clintons-convicted-child-abductor/ |
| mediaite.com | https://www.mediaite.com/tv/nbc-news-pete-williams-reports-fbi-really-isnt-conducting-clinton-foundation-investigation/ |
| mobile.wnd.com | http://mobile.wnd.com/2016/11/obama-claims-presidents-have-power-to-violate-constitution/#xhmqgbuwzzr6y9wd.99 |
| mostdamagingwikileaks.com | http://www.mostdamagingwikileaks.com/ |
| mrctv.org | http://www.mrctv.org/blog/flashback-jesse-jackson-praises-trump-seriousness-and-commitment-diversity-project |
| nationalreview.com | http://www.nationalreview.com/corner/442059/dont-blame-clinton-trump-2016-wouldve-beaten-obama-2012 |
| naturalnews.com | http://www.naturalnews.com/055879_associated_press_eric_tucker_dishonest_journalism.html |
| newsbusters.org | http://www.newsbusters.org/blogs/nb/kyle-drennen/2016/11/03/nets-ignore-massive-bombshell-fbi-investigation-clinton-corruption |
| newsmax.com | http://www.newsmax.com/politics/trump-president-salary-refuse/2015/09/18/id/692155/ |
| nydailynews.com | http://www.nydailynews.com/news/national/calif-woman-accusing-trump-child-rape-break-silence-article-1.2855631 |
| occupydemocrats.com | http://occupydemocrats.com/2016/11/03/fox-news-just-admitted-made-story-hillary-indicted-foundation/ |
| pamelageller.com | http://pamelageller.com/2016/11/must-see-video-undercover-journalist-in-full-burka-is-offered-huma-abedins-ballot-voterfraud.html/ |
| pastebin.com | http://pastebin.com/36q0yksm |
| patdollard.com | http://www.patdollard.com/75-of-americans-say-media-biased-for-hillary/ |
| patriotpost.us | https://patriotpost.us/articles/45877 |
| politopinion.com | http://politopinion.com/2016/11/undercover-pv-offered-huma-ballot/ |
| puppetstringnews.com | http://www.puppetstringnews.com/blog/black-voters-for-trump-double-from-2012-election |
| rasmussenreports.com | http://www.rasmussenreports.com/public_content/politics/elections/election_2016/white_house_watch_nov3 |
| redstate.com | http://www.redstate.com/aglanon/2016/11/07/for-the-sake-of-the-republican-party-dont-vote-for-trump/ |
| redstatewatcher.com | http://redstatewatcher.com/article.asp?id=46969 |
| scooprocket.com | http://scooprocket.com/us/2016/11/01/fbi-dumps-documents-from-bill-clintons-pardon-of-tax-cheat/ |
| shareblue.com | http://shareblue.com/60-million-hillary-voters-will-not-be-silenced/ |
| silenceisconsent.net | http://silenceisconsent.net/explosive-proof-hillary-tied-to-child-trafficker-laura-silsby |
| stateofthenation2012.com | http://stateofthenation2012.com/?p=54773 |
| theamericanfirst.com | http://theamericanfirst.com/video-cnn-interviews-leftist-protester-later-revealed-as-cnn-cameraman/ |
| theamericanmirror.com | http://www.theamericanmirror.com/obama-talks-207-times-campaigning-hillary/ |
| theblacksphere.net | http://theblacksphere.net/2016/11/michelle-obama-deleted-hillary-from-twitter/ |
| theconservativetreehouse.com | https://theconservativetreehouse.com/2016/11/01/epic-rick-santorum-blasts-nevertrump-and-john-kasich-video/ |
| thefederalist.com | http://thefederalist.com/2016/11/04/im-voting-trump-street-fighter-can-take-clinton-machine/#.wb4y8g-asaq.twitter |
| thefederalistpapers.org | http://thefederalistpapers.org/?p=100260 |

| Domain Name | Example URL |
| --- | --- |
| thegatewaypundit.com | http://www.thegatewaypundit.com/2016/11/donald-trump-takes-nearly-6-point-lead-crookedhillary-latest-la-times-poll/ |
| theodysseyonline.com | https://www.theodysseyonline.com/calling-millennials-lets-break-party-system |
| thepoliticalinsider.com | http://www.thepoliticalinsider.com/police-sources-confirm-officials-charges-bill-clinton-pedophile/ |
| therealstrategy.com | https://therealstrategy.com/wikileaks-admits-source-dnc-leaks-murdered-clintons/ |
| therebel.media | https://www.therebel.media/congratulations_america_you_won_your_second_war_of_independence |
| truepundit.com | http://truepundit.com/breaking-bombshell-nypd-blows-whistle-on-new-hillary-emails-money-laundering-sex-crimes-with-children-child-exploitation-pay-to-play-perjury/ |
| truthfeed.com | http://truthfeed.com/breaking-new-wikileak-bombshell-shows-cnn-asked-the-dnc-what-questions-to-ask-trump/34331/ |
| ukok.page.tl | http://ukok.page.tl/trump-train.htm |
| usalibertynews.com | http://usalibertynews.com/breaking-la-times-reporter-fired-tweeting-wants-trump-dead/ |
| vaskal.ca | http://www.vaskal.ca/podestafiles |
| weaselzippers.us | https://www.weaselzippers.us/305882-fbi-agents-are-now-talking-say-comey-stood-in-the-way-of-clinton-email-investigation/ |
| westernjournalism.com | https://www.westernjournalism.com/lawsuit-filed-against-attorney-general-loretta-lynch-over-tarmac-meeting-with-bill-clinton/?utm_source=twitter&utm_medium=mobilefloatingsharingbuttons&utm_content=2016-11-02&utm_campaign=websitesharingbuttons |
| wnd.com | http://www.wnd.com/2016/10/clintons-black-son-to-make-bombshell-announcement/ |
| youngcons.com | http://www.youngcons.com/clinton-foundation-doctored-memo-to-cover-distributing-watered-down-aids-medicine-in-africa/ |
| yournewswire.com | http://yournewswire.com/nypd-hillary-clinton-child-sex-scandal/ |